# Dynamical generalizations of the Drake equation: the linear and non-linear theories

## A.D. Panov

Moscow State University, Russia

**Abstract.** The Drake equation pertains to the essentially equilibrium situation in a population of communicative civilizations (CCs) of the Galaxy, but it does not describe dynamical processes which can occur in it. Both linear and non-linear dynamical population analysis is build out and discussed instead of the Drake equation.

**Introduction**

The crucial question of the SETI problem is how far the nearest CC from us is. Its answer depends on the number of CCs existing in the Galaxy at present. Fig. 1 shows how the distance between the Sun and the nearest CC depends on the number of CCs in the Galaxy. The calculation was fulfilled by the Monte Carlo method with the use of a realistic model of the distribution of stars in the Galaxy [1] and for the actual location of the Sun in the Galaxy (8.5 kpc from the center).

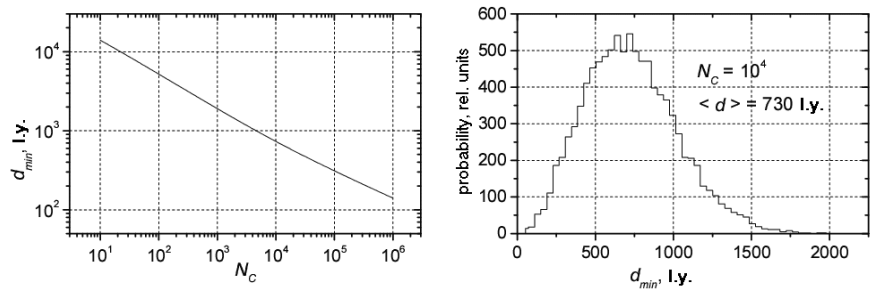

Fig. 1. The expected distance to the nearest CC as a function of the number of CCs in the Galaxy (left panel) and the probability distribution of distances to the nearest CC for the case of $N_C = 10000$ (right panel). The distribution function profile for other values of $N_C$ is analogous; only the most probable distance differs.

The best known way to answer the question about the number of CCs is the formula by F. Drake

$$N_C = R_* f_p n_e f_l f_i f_c L, \qquad (1)$$

where $R_*$ is a star-formation rate in the Galaxy averaged with respect to all time of its existence, $f_p$ is the part of stars with planet systems, $n_e$ is the average number of planets in systems suitable for life, $f_l$ is the part of planet on which life did appear, $f_i$ is the part of planets on which life developed to intelligent forms, $f_c$ is the part of planets on which life reached the communicative phase, $L$ is the average duration of the communicative phase. The Drake formula gives the number of CCs only in a rather rough approximation. According to the formula, $N_C$ does not depend on time. Meanwhile, it is evident that formerly there were no CCs in the Galaxy at all. Then there was a transition period when its number was increasing somehow. In fact, the Drake formula describes only the essentially stable situation, which can be very remote from the truth.

It is necessary to modify the Drake formula to allow for the development times, the variability of star formation rate, etc. Our paper develops this approach both in linear and non-linear dynamical theory.

**Linear population analysis**

In the linear theory it is supposed that the CCs develop independently from each other and that CCs cannot effect the star formation rate and evolution of life on other planets in the Galaxy. The following model functions and parameters are used in the model. $R(M,T)$ is the star formation rate as a function of star mass $M$ and galactic time $T$. The star lifetime is determined by the survival probability $L_S(M,\tau)$ of the star mass $M$ on the Main Sequence at the moment $\tau$ reckoned from the moment of its birth. $B(M,\tau)$, determines the density of the probability that a CC appears in the time $\tau$ after formation of a star of the mass $M$. The function $B(M,\tau)$ is normalized by $\int B(M,\tau)d\tau = \alpha(M)$, where $\alpha(M)$ gives a probability that conditions suitable for origin of a CC near the star of the mass $M$ will be implemented someday having infinitely long lifetime for the parent star. The duration of the communicative phase of CC evolution is determined by the function $L_C(M,\omega)$ that gives the probability of maintenance of the communicative phase in the time $\omega$ after its origin. The population of stars is described by distribution $n_S(M,\tau,T)$ specifying the number of stars by their mass $M$ and age $\tau$ with the galactic time $T$. The population of CCs is described by distribution $n_C(M,\tau,\omega,T)$ specifying the number of communicative civilizations of the age $\omega$ at the galactic time $T$ which appeared near the star having the mass $M$ and the age $\tau$. The total number of civilizations $N_C$ is:

$$N_C(T) = \int_0^\infty dM \int_0^T d\tau \int_0^{T-\tau} d\omega \, n_C(M,\tau,\omega,T). \qquad (2)$$

The complete system of equations together with margin conditions that determines the distributions $n_S(M,\tau,T)$ and $n_C(M,\tau,\omega,T)$ is

$$\frac{\partial n_S}{\partial T} = -\frac{\partial n_S}{\partial \tau} - \Lambda_S(M,\tau)n_S; \quad -\Lambda_S(M,\tau) \stackrel{def}{=} \frac{\partial \ln L_S(M,\tau)}{\partial \tau} \qquad (3)$$

$$n_S(M,\tau,0) = 0 \qquad (4)$$

$$n_S(M,0,T) = R(M,T) \qquad (5)$$

$$\frac{\partial n_C}{\partial T} = -\frac{\partial n_C}{\partial \omega} - [\Lambda_C(M,\omega) + \Lambda_S(M,\tau+\omega)]n_C \qquad (6)$$

$$n_C(M,\tau,\omega,0) = 0 \qquad (7)$$

$$n_C(M,\tau,0,T) = n_S(M,\tau,T)B(M,\tau). \qquad (8)$$

The definition of $\Lambda_C$ in (6) is obvious from (3).. The system (3-8) has exact solution for the density distribution of CCs as follows:

$$n_C(M,\tau,\omega,T) = R(M,T-\tau-\omega)L_S(M,\tau+\omega)B(M,\tau)L_C(M,\omega). \qquad (9)$$

The obtained solution (9) together with formula (2) allows us to investigate a huge number of various tasks. It is worth to restrict this variety by some reasonable limits. For this purpose, some simplifications will be used. We suppose additionally: the star formation rate may be factorized: $R(M,T) = R_*(T)F(M)$, where the initial mass function $F(M)$ is supposed to be independent of time; the star survival probability to be a step-like function: $L_S(M,\tau) = \Theta[\tau_0(M) - \tau]$, where $\tau_0(M)$ is the lifetime of the star with mass $M$ on the Main Sequence; the time of development before CC formation to be independent of the star mass $M$: $B(M,\tau) = \alpha(M)b(\tau)$ with $\int b(\tau)d\tau = 1$; the lifetime of CC does not depend on the star mass $M$: $L_C(M,\omega) = L_C(\omega)$. Expression (2) with using of (9) and the introduced simplified expressions for the model functions may be rewritten as

$$N_C = \int_0^\infty dM \, \alpha(M)F(M) \int_0^T d\tau \, b(\tau) \int_0^{\omega_{max}(M)} d\omega R_*(T-\tau-\omega)L_C(\omega), \qquad (10)$$

where $\omega_{max}(M) = \min[T-\tau, \tau_0(M)-\tau]$.

The Drake equation (1) may be obtained from eq. (10) with further simplifications: $R_*$=const, $\tau_0(M) \equiv \infty$, time of development before CC formation is small. But we will investigate more realistic scenarios.

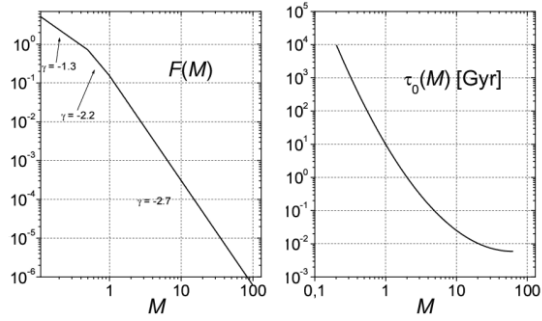

Fig. 2. *Left panel*: The initial spectrum of star masses. *Right panel:* The star lifetimes. The star mass is in the solar mass. The quantity γ in the diagram of *F*(*M*) shows an index of the power function corresponding to different parts of the spectrum

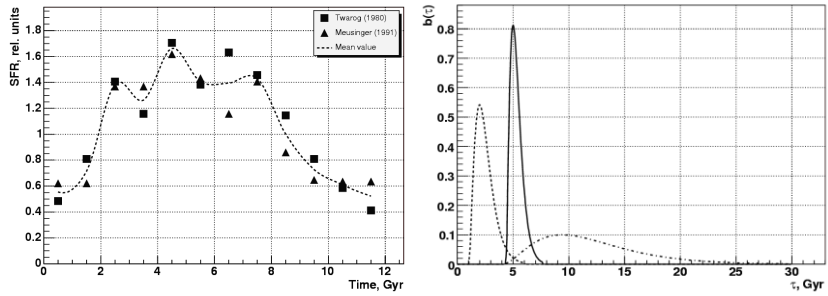

Fig. 3. Left panel: star formation rate as a function of time. Right panel: the choice of probability of CC development times. Solid line represents the most probable case.

The initial spectrum of star masses according to [2] and the relation between star lifetimes on the Main Sequence and mass approximated in the [3, p. 58] were used in calculations. Fig. 2 shows corresponding functions $F(M)$ and $\tau_0(M)$. For the star formation rate function $R_*(T)$ in calculations we used averaged and interpolated data from the papers of [4] and [5] (shown by the dotted line in Fig. 3, left panel). The relative rate data of [3,4] were normalized to obtain correct number of stars in the Galaxy at the present time. The linear function equal to zero at $M = 0.5 M_\odot$, equal to 1 at $M = 2 M_\odot$ and $\alpha(M) = 1$ at $M > 2 M_\odot$ was taken as the probability of realization of suitable conditions. The value $\alpha(M) = 1$ for $2 M_\odot$ was chosen rather arbitrarily and it does not restrict the generality due to linearity of the theory. With such choice of $\alpha(M)$ the average probability of implementation of suitable conditions with star masses from $0.5 M_\odot$ to $2 M_\odot$ turns out to be about 0.02. For the distribution density of CC development times $b(\tau)$ we tested here three func-tions shown in Fig. 3, right panel. The distribution of durations of the communicative phase was taken in the form of the falling exponent $L_C(\omega) = \exp(-\omega/L_0)$ with $L_0 = 1000$ years. The choice of $L_0$ practically does not limit the generality of results (due to linearity).

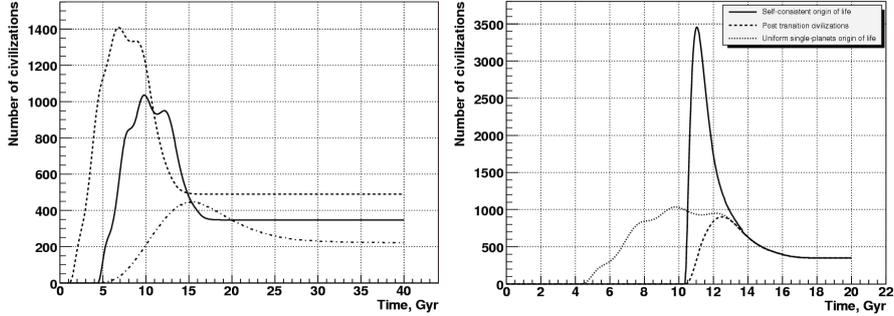

Fig. 4. *Left panel:* results of calculations within the framework of the simple linear theory corresponding to different distributions $b(\tau)$ of CC development time (see Fig. 3). *Right panel:* The linear dynamics of the CC population at the origin of life in the Galaxy in the process of the self-consistent phase transition [6] in 6 billion years after the beginning of the formation of the galactic disk and its comparison with the simple linear dynamics at the constant formation of CCs with the development time 5 billion years (for references: actual age of the Galaxy disc is 12 billion years).

Fig. 4 (left panel) shows results of calculations carried out with the above assumptions with the formulae (9,10). The results correspond to different distributions of CC development times. All curves have a strongly pronounced maximum associated with an SFR peak at T≈5 billion years (see Fig. 3). The peak in the number of civilizations is a linear response to it and can be called *a linear demographic wave*. For the basic variant of calculations (the solid line) the present time (12 billion years) falls within the region of the maximum of the linear demographic wave.

Note that though the relations in Fig. 4 (left panel) are constructed for a very limited set of parameters, they can be used to estimate within the context of many other scenarios. So, the curve amplitude will be proportional to the average CC lifetime (the parameter $L_0$) and the curve amplitude will also be proportional to the maximum probability of realization of suitable conditions (the maximal value in $\alpha(M)$ function, see above).

Up to this point the conditions leading to the origin of a CC have been supposed to be unchanged during the history of the galactic disk. Actually, variations of them are possible for a number of reasons (variable background of cosmic rays, etc). The conditions change for sure if the hypothesis about the self-consistent galactic origin of life and related phase transition [6] is true. In this case a great "Big Bang" of life origin took place in the history of the Galaxy and if the development time to CC is more or less standard (like $b(\tau)$ presented by solid line in Fig. 3, left panel – about 5 billion years), then "Big Bang" of CCs origins should be followed as well. The theory describing this *phase demographic peak* may be deduced from the described above linear theory (we omit the details) and the results of calculation are presented in Fig. 4 (right panel). It was supposed the "Big Bang" of life origin to be 6 billion years after the start of formation of the Galaxy disk and the average time of development for CC to be 5 billion years in this calculation. The dashed line in Fig. 4 (right panel) shows the partial distribution for planets with the origin of life after the "Big Bang" of life origin (as Earth). One can see that we can live both before and after the phase peak.

Thus, the population analysis based on the linear theory and real astrophysical data predicts no-trivial dynamical patterns of evolution of CCs like the linear demographic wave and the phase peak (Fig. 3). Non-linear generalization of the formalisms leads to even more interesting picture.

**Non-linear population analysis**

In the linear theory given above the distributions $B(M, \tau)$ and $L_C(\omega)$ describing the origin and life of communicative civilizations were supposed to be independent on the number of available civilizations. The function $R(M,T)$ describing the "natural" star formation rate was also considered to be independent of the CC population. This is true until civilizations have no effect on one another, nor on conditions of origin of other civilizations, nor on conditions of origin of stars. The theory accounting for this effect ceases to be linear.

The first possibility for a non-linear theory is related to the influence on the function $R(M, T)$ – "the artificial creation of stars". The second possibility – the influence on the distribution $B(M, \tau)$ – must imply some sort of directed panspermia of life or intelligent life. The third kind of non-linear phenomena related to the changing of the probability $L_C(\omega)$ by a mutual influence of civilizations through contacts by communication channels. We thoroughly study only the last possibility here. Other options may be studied by similar methods.

Without limiting generality, the CCs may be thought to be divided into three categories: the CCs for which the contact is "harmful", because it reduces the duration of the communicative phase, the neutral CCs, and the CCs for which the contact is "useful", because it prolongs the communicative phase. We will call the last category *extrovert* civilizations and will denote them as ECC. In the following we will consider the dynamics of the subpopulation of the extrovert civilizations only.

It can be supposed additionally that one of the most important properties of ECCs is an increase in efficiency of search for partners and establishment of communication under the influence of the already established contacts (we call it *civilization range*). This circumstance will be substantially used below.

It is important that if ECCs do exist, then a process with a positive feedback can begin. The larger is the number of ECCs in the Galaxy, the higher is the contact probability. The contact increases the lifetime of the ECC and its civilization range, which leads to increasing the ECC population, which rises the contact probability again, and so on. The positive feedback loop can lead to an avalanche-like phase transition in the Galaxy-scale accompanied by a powerful burst of the number of ECCs. ECCs become prevailing in the Galaxy even if the situation was different before the transition. Some details of this phenomenon are described by the formalism proposed below.

In the linear theory the current state of a separate civilization was described only by age of the communicative phase $\omega$, which, in combination with the star lifetime and the moment of the civilization origin, made it possible to statistically predict the fate of a civilization. To account for the mutual influence through communication channels they should be described in greater details. We will consider that every civilization is described by age $\omega$ and by a vector of parameters $\mathbf{q}$ that will be called "a quality". This is a set of characteristics of ECCs which affects, first of all, an expected duration of the communicative phase and civilization range. It is supposed that the contact increases the ECC quality in a sense, and thanks to that the communicative phase prolongs and civilization range increases. Thus, the probability of civilization survival should be considered as dependent on its quality which must be also one of the arguments of the civilization distribution function:

$$L_C(M,\omega) \to L_C(M,\mathbf{q},\omega); \quad n_C(M,\tau,\omega,T) \to n_C(M,\tau,\mathbf{q},\omega,T).$$

To describe the influence of contacts of a civilization $A$ with a number of other civilizations $B_1, B_2,\ldots$ on the quality of $A$ we suppose the effect to be additive:

$$\frac{d\mathbf{q}_A}{dt} = \sum_i K(\mathbf{q}_A, \omega_A, \mathbf{q}_{B_i}, \omega_{B_i}), \tag{11}$$

where $K(\mathbf{q}_A, \omega_A, \mathbf{q}_B, \omega_B)$ is some universal function representing the contact model. Obviously the additive model of contacts is a simplification that may be reasonable only in a case of low number of contacts per civilization.
Equations (3–5) for the star distribution function and equations (6–8) for the civilization distribution function remain valid in non-linear dynamics. Only a new term appears in it describing "the current" of the civilization quality in the *q*-space due to interaction between them. Besides, now the edge condition must describe weights of ECCs quality starting the communicative phase. The total system of equations for the distribution function $n_C(M,\tau,\mathbf{q},\omega,T)$ is written in the following way:

$$\frac{\partial n_C}{\partial T} = -\frac{\partial n_C}{\partial \omega} - [\Lambda_C(M,\mathbf{q},\omega) + \Lambda_S(M,\tau+\omega)]n_C$$
$$-\nabla_\mathbf{q}[j(\mathbf{q},\omega,T)n_C] \qquad (12)$$

$$n_C(M,\tau,\mathbf{q},0,T) = 0 \qquad (13)$$

$$n_C(M,\tau,\mathbf{q},0,T) = n_S(M,\tau,T)B(M,\tau,\mathbf{q}) \qquad (14)$$

The problem of calculation of the **q**-current $j(\mathbf{q},\omega,T)$ generally is very difficult but it may be solved for the additive model of contacts (11) and for the model of a large homogeneous galaxy (edge effects can be neglected):

$$j(\mathbf{q},\omega,T) = \frac{4\pi c^3}{V_G}\int d\omega'\int d\mathbf{q}' K(\mathbf{q},\omega,\mathbf{q}',\omega')\int_{T-r(\mathbf{q},\omega,\mathbf{q}',\omega')/c}^{T} dT'(T-T')^2 \times$$
$$\int dM \int d\tau\, n_C(M,\tau,\mathbf{q}',\omega',T') \qquad (15)$$

In formula (15) $V_G$ is the galaxy volume, $c$ is the velocity of light, and $r(\mathbf{q},\omega,\mathbf{q}',\omega')$ is the range of communication between two civilizations with the qualities and the ages $(\mathbf{q},\omega)$ and $(\mathbf{q}',\omega')$.

Due to the term corresponding to the quality current, (12) turns out to be very complicated integro-differential equation. However, it may be solved numerically for simple models of contact under some additional simplifying assumptions as described below.

The civilization quality will be considered to be presented by the only scalar parameter $q$. It is supposed the average value of the quality for an isolated civilization (without any contacts) is $q = 1$. We transfer from the detailed description of a civilization by its quality and age to the average value of quality upon the whole lifetime of the civilization and averaged upon all star masses. Further, we consider the number of civilizations per unit of volume of a uniform galaxy. That is, instead of the exact distribution $n_C(M,\tau,q,\omega,T)$ we consider averaged distribution $\rho(q,T)$ such that $V_G\int\rho(q,T)dq = N_C(T)$. We consider the civilization origin rate normalized per galaxy volume unit to be a given function of the galaxy time $f(T)$, and the distribution density of the parameter $q$ for isolated civilizations to be $\phi_o(q)$ such that $\int\phi_o(q)dq = 1$ and the mean value of $\phi_o(q)$ is equal to 1. Then equations (12–14) can be transcribed in the form of a single equation

$$\frac{\partial\rho(q,T)}{\partial T} = -\Lambda_C(q)\rho(q,T) + f(T)\phi_0(q) - \frac{\partial}{\partial q}[j(q,T)\rho(q,T)]. \qquad (16)$$

Initial conditions for the function $\rho(q,T)$ can be specified at any time $T = T_0$, and equation (16) can be solved as the Cauchy initial-value problem.

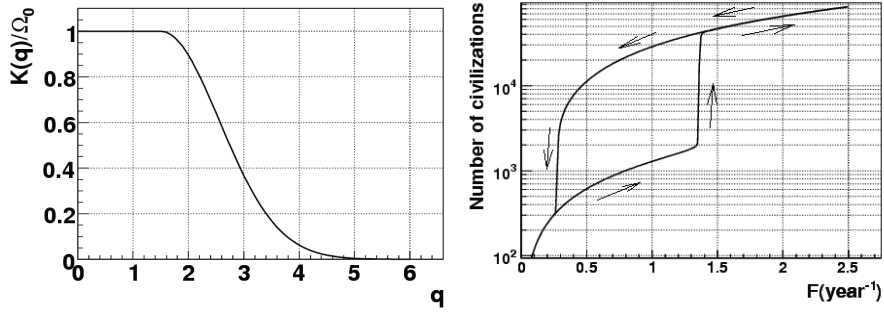

Fig. 5. *Left panel:* The function $k(q)$ used in calculations. *Right panel:* The bistability in a ECC population obtained by numerical solution of the equation (16).

To calculate the divergence term in (16) we adopted the following simple model of contact

$$\frac{dq_A}{df} = k(q_A) q_A \sum_i q_{B_i} ,\qquad(17)$$

where the function $k(q)$ is shown in Fig. 5 (left panel) with $\Omega_0 = 0.001$ years$^{-1}$. With (17) the equation (15) is simplified to

$$j(q,T) = 4\pi c^3 q k(q) \int dq' q' \int_{T-r(q,q')/c}^{T} dT' (T-T')^2 \rho(q',T') .\qquad(18)$$

For the inverse lifetime function $\Lambda_C(q)$ and the range function $r(q,q')$ various assumptions may be taken but we adopted the following ones here:

$$\Lambda_C(q) = \Omega_0 / q^2, \quad \Omega_0 = 0.001\ years^{-1} \qquad(20)$$

$$r(q_A, q_B) = r_0 \times (q_A q_B)^{1/5}, \quad r_0 = 400\ l.y. \qquad(21)$$

The expression for $r(q_A,q_B)$ was obtained under the assumption that reception and transmission are fulfilled only by a beam antenna (we have to omit the details of the explanation). The distribution $\phi_0(q)$ was taken to be gaussian one with mean value 1 and dispersion 0.2.

Some results of calculations are shown in Fig. 5 (right panel). Let us elucidate the computing technique and sense of the obtained results. It was supposed that at the initial time $T = 0$ there were no civilizations, $\rho(q,0) = 0$. After that the civilization origin rate $F$ begins increasing slowly, so that at any time an almost complete equilibrium is achieved in the population of ECC. Fig. 5 (right panel) shows the relation between the number of civiliza-tions and $F$ (both normalized to the volume of our Galaxy). The equilibrium number of civilizations increases as $F$ increases. In the process, first a point in the diagram moves along the lower branch of hysteresis loop from left to right and the number of civilizations is still small (less than 2000). This is the *silence epoch*, the probability $P$ to find a partner to contact for any civilization is $P << 1$.

However, due to the increasing number of civilizations, the situation becomes unstable, and when $F$ achieves a value of about 1.35 civilizations per year, and $P \approx 0.05$, then the equilibrium is broken. Due to the positive feedback between the number of contacts, civilization ranges and lifetimes, the number of civilizations and the probability of their interaction start increasing as an avalanche. As this takes place, the number of civilization increases sharply by about an order, and the average number of partners per one civilization achieves 10. This phase transition ends because the possibility of "improving" is exhausted at large values of the quality $q$ (see Fig. 5, left panel). The *saturation of contacts epoch* starts ($F > 1.4$).

Then, in calculation, the civilization origin rate stops increasing (at $F = 2.5$ yeas$^{-1}$) and begins the slow. First, a point in the diagram moves backwards, repeating the trajectory of $F$ growth. However, when reaching a critical value of $F = 1.35$ per year the reverse transition does not occur. This is prevented by the positive feedback "number of contacts – lifetime and range". The contact saturation epoch continues. Here two different stable states of the civilization population correspond to every value of $F$: one on the lower branch of hysteresis loop, the other on the upper branch. This is the bistability phenomenon. Only when $P$ approaches a value of about 0.5, the positive feedback already cannot keep the contact saturation phase from destruction, the number of civilizations sharply fall, and the silence epoch returns.

We neglected by fluctuations of density of civilization, but fluctuations can create the saturation of contacts phase locally with subsequent growth.